\documentclass[11pt,twoside]{article}

\usepackage{asp2004}
\usepackage{epsf}
\usepackage{psfig}
\usepackage{lscape}

\begin{document}

\title{Exploring uncharted territory in particles physics using pulsating white
dwarfs: Prospects}
\author{Agnes Kim, D.E. Winget, M.H. Montgomery}
\affil{The University of Texas at Austin, Astronomy Department, 1 University
Station, C1400, Austin, TX 78712, USA}
\author{and}
\author{D.J. Sullivan}
\affil{School of Chemical and Physical Sciences, Victoria, and University of
Wellington, P. O. Box 600, Wellington, New Zealand}

\begin{abstract}

Pulsating white dwarfs, especially DBVs, can be used as laboratories to study
elusive particles such as plasmon neutrinos and axions. In the degenerate
interiors of DBVs, plasmon decay is the dominant neutrino producing process. We
can measure the neutrino luminosity using asteroseismology and constrain plasmon
neutrino rates. In the same way, we can measure any additional loss of energy
due to other weakly interacting particles, such as axions. Depending upon their
(theoretically largely unconstrained) mass, axions could be a significant source
of energy loss for DAVs as well. We are looking at what the uncertainties in the
observables are, and what mass and temperature range minimizes them.

\end{abstract}

\section{Introduction}

Above a temperature of about 26000 K, more than half of the luminosity of a
white dwarf comes from neutrino emission (Fig. 1). In that temperature range,
the neutrinos are mainly produced by the decay of plasmons. So if we measure a
neutrino luminosity in hot white dwarfs, we are measuring plasmon neutrino
rates. O'Brien \& Kawaler (2000) tried to apply this idea to pulsating pre-white dwarf
stars, which have the highest neutrino luminosities. Because of difficulties
involved in modeling pre-white dwarf stars, they were not able to disentangle
the effect of neutrino energy loss from the effect of the contraction of those
stars.

Following the discovery of pulsating DBs as hot as 28000K, Winget et al. (2004)
demonstrated that one could use hot DBVs to measure neutrino rates, without
having to worry about contraction. At these temperatures, white dwarfs have
cooled to a point where they have become nearly fully degenerate and
contraction no longer has a large effect on the rate of change of pulsation
periods (\.{P}).

One can use the change in the pulsation periods over time to measure the
neutrino luminosity. As a white dwarf cools, the period of a given mode
increases. The faster the cooling, the faster the period increases. Mestel
theory (Mestel 1952) predicts \.{P} if the white dwarf is leaking energy
exclusively through photons. A higher \.{P} than expected means that the star
is cooling faster than expected, and indicates an extra source of energy loss.
\. {P} provides therefore a measure of the neutrino luminosity.

Axions, if they exist, would have an effect qualitatively similar to neutrinos,
in the sense that they too would escape the interior without interacting very
much and constitute an extra source of energy loss. An axion luminosity may be
determined from the rate of period change in the same way. One may ask at this
point how to tell apart the contributions from neutrinos and axions. We address
this question in section 3.2.

\begin{figure}
\center{
\leavevmode
\epsfxsize=75mm \epsfbox{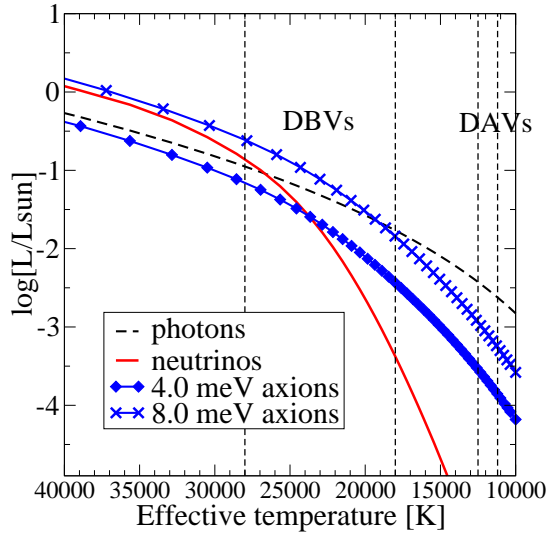}
\caption{Time evolution of different sources of energy loss for a 0.59 $M_\odot$
fiducial model. The vertical dashed lines mark the DBV and DAV instability
strips. The neutrino luminosity remains significant near the blue edge of the
DBV instability strip.}
\label{Figure 1}
}
\end{figure}

\section{Plasmon neutrinos}

In this section, we give a brief summary of the physics of plasmon neutrinos.
For a more detailed treatment, we recommend Winget et al. (2004) and references
therein. Plasmon neutrinos result from the decay of a plasmon into a
neutrino-antineutrino pair. A plasmon of frequency $\omega$ is made up of an
oscillating electromagnetic field coupled with electrons oscillating with the
same frequency. Classically, one thinks of a plasmon as an electromagnetic wave
propagating through a dielectric medium. The frequency of such a wave obeys the
dispersion relation $$\omega^{2}=\omega_{\rm o}^{2}+k^2c^2.$$ Quantizing the
field, the equation above gives $$E^2=p^2c^2+\hbar^2\omega_{\rm o}^2.$$ In
effect, a plasmon is a particle similar to a photon, except it has a non-zero
rest mass.

In free space, photons cannot decay into a neutrino-antineutrino pair without
violating conservation of four-momentum. In a plasma, the electrons coupled to
the photon allow conservation of energy and momentum and so plasmons can decay
into a neutrino-antineutrino pair.

\subsection{Detectability in DBVs}

To constrain the neutrino luminosity, one needs to make a certain number of
observations and construct a good model. The observables are the mass, the
effective temperature, the pulsation spectrum, and \.{P}. In order to use that
information to constrain the physics under study (i.e. the plasmon neutrino
rates), we need to make sure that we input the correct mass and effective
temperature. Uncertainties in the stellar parameters lead to uncertainties in
the plasmon neutrino rates. There are other sources of uncertainties to explore
as well, such as surface layer masses and core composition, but the present work
focuses on the effects of uncertainties in mass and effective temperature.

Figure 2 shows the expected \.{P} for different k modes of an l=1 mode of a
GD358-like fiducial model (in other words, a typical hot DBV). To obtain the
error bars, we ran a grid of models around the fiducial model differing in
effective temperature by $\pm$5\% and in mass by $\pm$10\%. For each model, we
ploted \.{P} versus k, and considered the largest departure from the fiducial
model's curve.

We used the analytical fits to the neutrino rates from Itoh et al. (1996). It is
clear from the figure that if plasmon neutrinos do exist and the actual rates
are accurately predicted by theory, then we should see a clear signature in the
\.{P}s. Quantitatively, the above uncertainties in the mass and effective
temperature translate to uncertainties in neutrino luminosity of about 50\%.

We have established that hot DBVs offer us a way to place meaningful constraints
on plasmon neutrino rates. To do this in practice, we need to find stable
pulsating DBs and measure their \.{P}. So far, we have one promising candidate,
EC20058 (Koen et al. 1995; Sullivan \& Sullivan 2000). With our current instruments, we could
probably get a \.{P} for EC20058 with a 5 year baseline.

\begin{figure}[t]
\center{
\leavevmode
\epsfxsize=80mm \epsfbox{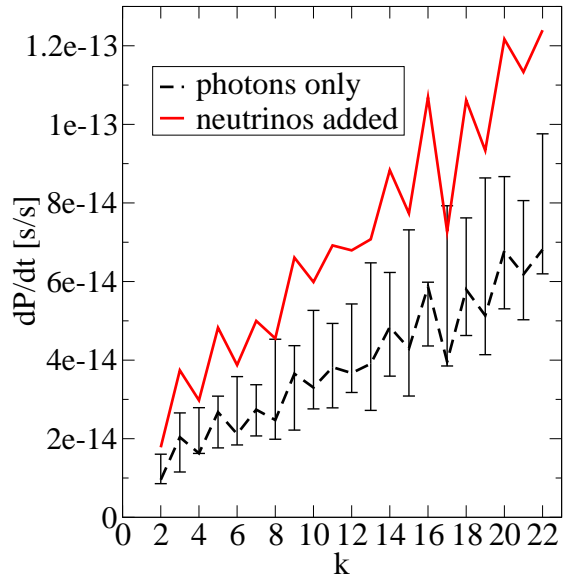}
\caption{Expected \.{P} for different k modes of an l=1 mode of a GD358-like
fiducial model (0.65 $M_\odot$, 22600 K)}
\label{Figure 2}
}
\end{figure}

\section{Axions}

Axions were 'invented' to solve the strong charge-parity violation in Quantum
Chromodynamics. According to their quark structure, neutrons should have a
measurable dipole moment, and they do not. The problem is solved if one
introduces a new symmetry, the breaking of which gives rise to the axion. The
question of the existence of axions is made more important since they are one
of the current best candidates for dark matter.

In the DFSZ model (Dine et al. 1981), the axion luminosity in a white dwarf is given
by $$\epsilon_{ax}=1.08\times10^{23}\, {\rm \mbox{ergs} \; g^{-1}
s^{-1}}\;\frac{g_{ae}^2}{4\pi}\;\frac{Z^2}{A}\:T_7^4\;F$$
where $T_7$ is in units of $10^7K$ and F is a function of the local density and
temperature. Nakagawa et al. (1988) give analytical fits for F. For much of the
interiors of white dwarfs in the range of temperatures considered in this
paper, F is of order unity. $g_{ae}$ represents the strength of the coupling of
the axions with the electrons, on which the mass of the axion depends
($g_{ae}\propto m_a\cos^2\beta$). By measuring the axion luminosity, we can
constrain $g_{ae}$. In the astronomical community, $g_{ae}$ is commonly called
'axion mass', and we shall adopt this terminology here.

\subsection{Mass limits}
The lowest limit cosmology places on the mass of the axion is currently around
$10^{-5}$ eV (Raffelt 1990). A number of astrophysical constraints and
particle physics experiments place upper limits on the axion mass. Microwave
cavity experirments probe narrow ranges in the $\mu{\rm {eV}}$ domain (see for instance Bradley et al. 2003). They have been unsuccessful so far, but are very
slowly eliminating possible ranges for the axion mass. Solar axion telescopes
are sensitive to axions as small as 1 eV in mass (Irastorza et al. 2003). Telescope
searches, which look for a characteristic emission line from galactic clusters
have a similar sensitivity to axions (Raffelt 1990). So far, both types of
searches have failed to detect any axions, and place an upper limit on the
axion mass of $\sim\: 1 {\rm {eV}}$.

The tightest constraints, however, come from astrophysical observations. One is
based on the arrival times of neutrinos from SN1987A. Neutrinos from SN1987A
provide information on the cooling of the exploding core. Axions more massive
than $\sim\,10$ meV would have accelerated the cooling in a way that was
inconsistent with the observations (Raffelt 1990) Another one is based on
horizontal branch stars. The lifetime of those stars would be shortened by a
loss of energy due to axions. By requiring that their lifetime be consistent
with the observed structure of horizontal branches in clusters, an upper limit
of $\sim\,10$ meV may be placed on the axion mass (Raffelt \& Weiss 1995).

Pulsating white dwarf stars offer a way to place an even tighter upper mass
limit. Figure 1 shows that unlike the neutrino luminosity, the axion luminosity
declines slowly. This means that axions may still play a role down into the DAV
temperature range, while neutrinos are not expected to. While we yet have to
measure a \.{P} for DBVs, we already have a solid measurement for G117-B15A
(Kepler et al. 2000). Using this result, Corsico et al. (2001) have placed an upper
limit on the axion mass of 4 meV.

\subsection{Detectability in DAVs and DBVs}

\begin{figure}[t]
\begin{center}
$\begin{array}{c@{\hspace{2mm}}c}
\epsfxsize=65mm
\epsffile{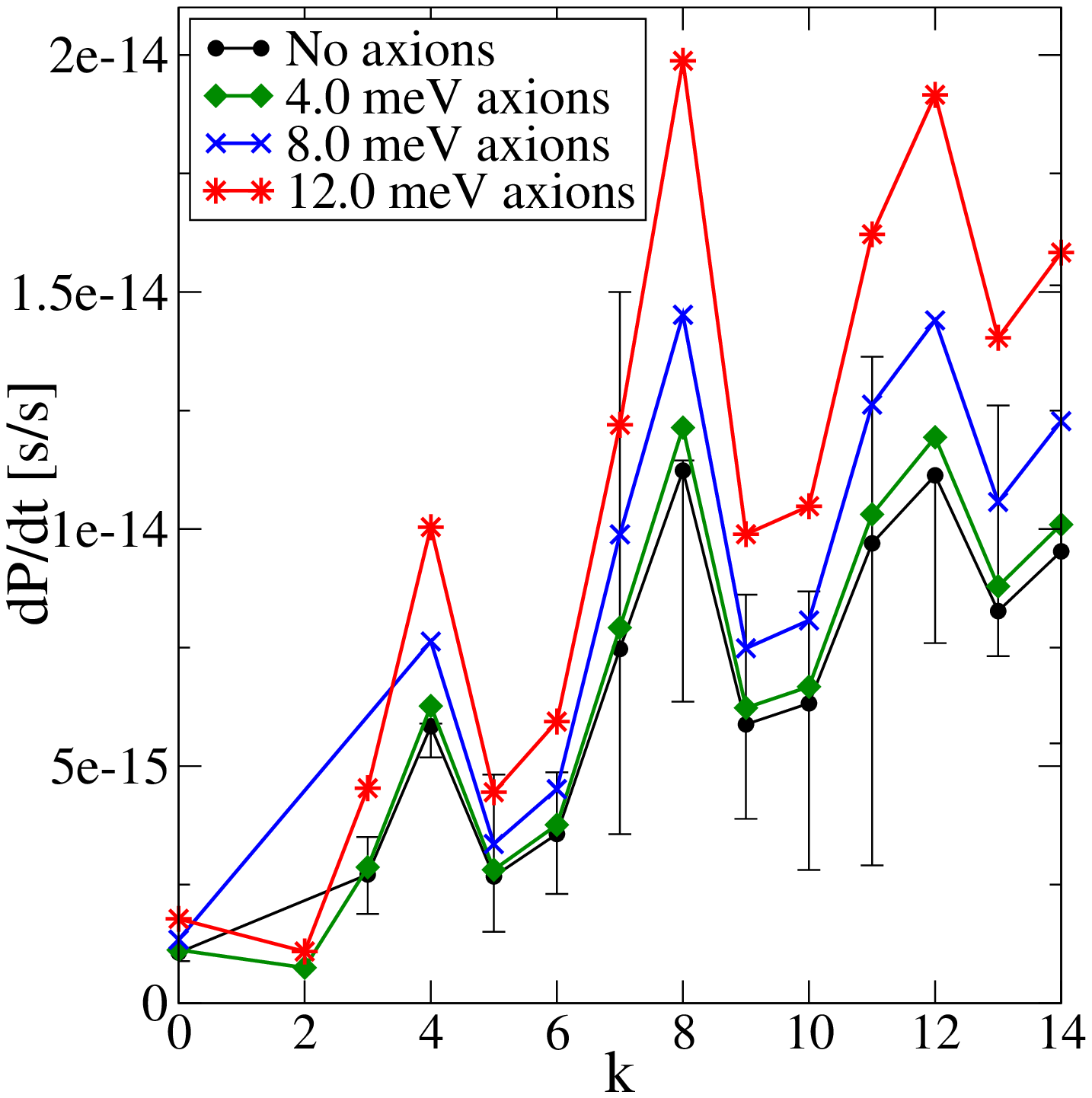} &
	\epsfxsize=65mm
	\epsffile{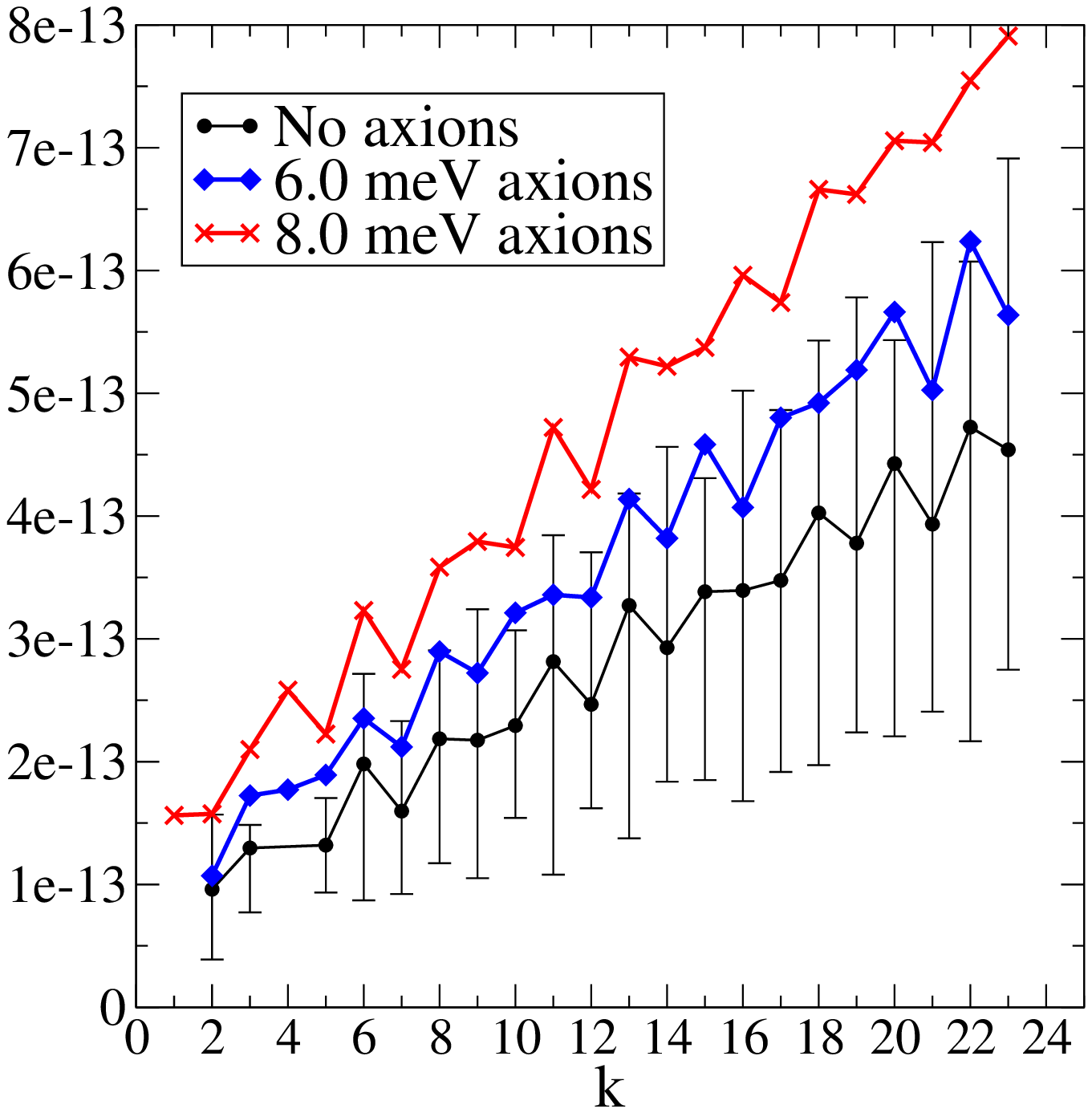} \\ [0.1cm]
\mbox{\bf (a)} & \mbox{\bf (b)}
\end{array}$
\end{center}
\caption{a) Expected \.{P} for different k modes of an l=1 mode of an R548-like
fiducial model (0.54 $M_\odot$, 11990 K) and for varying axion masses. b) Same
thing, for a hot DBV (0.60 $M_\odot$, 28000 K).}
\label{Figure 3}
\end{figure}

Figure 3a shows the expected \.{P} for different k modes of an $\ell=1$ mode of
an R548-like fiducial model (model parameters from Bradley 1998). The error
bars were obtained in the same way as in figure 2, except that the
uncertainties in mass and temperature considered were both $\pm$5\%. We used
analytical fits for the neutrino luminosity from Nakagawa et al (1988).

Considering the uncertainties,  axions could be as massive as $\sim\,8$ meV and
not have a clear effect on the measured \.{P}. That is, the difference between
the expected \.{P} and the measured one could be due entirely to the
uncertainties in mass and temperature, as opposed to an extra source of energy
loss. In that respect, the limit of 4 meV found by Corsico et al. (2001) may be
somewhat optimistic.

A tighter limit can be placed on the axion mass by looking at DBVs (Fig. 3b).
For a hot DBV (28000 K), 8 meV axions leave a clear signature. At that
temperature, however, neutrinos are still a significant source of energy loss
as well. How does one distinguish between the two? Winget et al. (2004)  have shown
that for a given effective temperature, the plasmon neutrino rates depended
strongly on white dwarf mass (for a given effective temperature). For instance,
according to predictions, massive DBs are no longer cooling through neutrinos.
Any additional source of energy loss for those stars around 28000 K would be
the result of axions.

\section{Conclusions}

Hot DBVs offer us a way to place meaningful constraints on plasmon neutrino
rates and a tighter upper limit on the axion mass, assuming we can distinguish
between the effects of neutrinos and axions. Looking at different mass DBVs is
one answer. A more immediate answer is to use DAVs.

Unlike neutrinos, axions (if massive enough) should still be a source of energy
loss for DAVs. Even though they do not allow us to place as tight a limit on
the axion mass, DAVs have two advantages. First, we already have a measurement
of \.{P} for one of them (G117-B15A). Second, they are no longer cooling
through neutrinos. For DAVs, theory tells us that any additional source of
energy loss can only come from axions, not neutrinos. This has allowed
Corsico et al. (2001) to place an upper mass limit of 4 meV. Even a more
conservative limit of $\sim\,8$ meV constrains the axion mass better than other
observations and experiments do.

The work was supported in part by ARP-0543 from the Texas Advanced Research
Program.

\end{document}